\documentclass{emulateapj}
\usepackage{graphics}

\shorttitle{Anti-parallel filament flows with Hi-C}
\shortauthors{Alexander et al.}

\begin{document}

   \title{Anti-parallel EUV flows observed along active region
     \\ filament threads with Hi-C} 
   \author{Caroline E. Alexander\altaffilmark{1}, Robert
     W. Walsh\altaffilmark{1}, St\'ephane R\'egnier\altaffilmark{1},
     Jonathan Cirtain \altaffilmark{2}, Amy
     R. Winebarger\altaffilmark{2}, Leon Golub \altaffilmark{3}, Ken Kobayashi\altaffilmark{4},
     Simon Platt\altaffilmark{5}, Nick Mitchell\altaffilmark{5},
     Kelly Korreck\altaffilmark{3}, Bart DePontieu\altaffilmark{6},
     Craig DeForest\altaffilmark{7}, Mark Weber\altaffilmark{3}, Alan
     Title\altaffilmark{6}, Sergey Kuzin\altaffilmark{8}} 
\vspace{0.5cm}
\affil{$^{1}$ Jeremiah Horrocks Institute, University of Central
     Lancashire, Preston, PR1 2HE, UK}
\affil{$^{2}$ NASA Marshall Space Flight Center, VP 62, Huntsville, AL
  35812, USA} 
\affil{$^{3}$ Harvard-Smithsonian Center for Astrophysics, 60 Garden
  St., Cambridge, MA 02138, USA} 
\affil{$^{4}$ The University of Alabama in Huntsville, Center for
  Space Plasma and Aeronomic Research, 320 Sparkman Dr, Huntsville, AL
  35805} 
\affil{$^{5}$ School of Computing, Engineering and Physical Sciences, University of Central
     Lancashire, Preston, PR1 2HE, UK} 
\affil{$^{6}$ Lockheed Martin Solar \& Astrophysics Lab, 3251 Hanover
  St., Org. ADBS, Bldg. 252, Palo Alto, CA} 
\affil{$^{7}$ Southwest Research Institute, 1050 Walnut Street, Suite
  300, Boulder, CO 80302} 
\affil{$^{8}$ P.N. Lebedev Physical institute of the Russian Academy
  of Sciences, Leninskii prospekt, 53, 119991, Moscow}

\begin{abstract}  
   
   {Plasma flows within prominences/filaments have been observed for many
years and hold valuable clues concerning the mass and energy balance
within these structures. Previous observations of these flows
primarily come from H$\alpha$ and cool EUV lines (e.g., 304\,\AA) where
estimates of the size of the prominence threads has been
limited by the resolution of the available instrumentation. Evidence
of `counter-steaming' flows has previously been inferred from these
cool plasma observations but now, for the first time, these flows have
been directly imaged along fundamental filament threads within the
million degree corona (at 193\,\AA). In this work we present observations of an active
region filament observed with Hi-C that exhibits anti-parallel flows
along adjacent filament threads. Complementary data from
SDO/AIA and HMI are presented. The ultra-high spatial and temporal 
resolution of Hi-C allow the anti-parallel flow velocities to be
measured (70\,-\,80\,km\,s$^{-1}$) and gives an indication of the resolvable
thickness of the individual strands (0.8''\,$\pm$\,0.1''). The temperature distribution
of the plasma flows was estimated to be log T\,(K) = 5.45\,$\pm$\,0.10
using EM loci analysis. We find that
SDO/AIA cannot clearly observe these anti-parallel flows nor measure their
velocity or thread width due to its larger pixel size. We suggest that
 anti-parallel/counter-streaming flows are likely commonplace within all 
filaments and are currently not observed in EUV due to current instrument
spatial resolution.}  
\end{abstract}


%________________________________________________________________

\section{Introduction}
\label{sec:intro}

Research into the mechanisms that create and maintain solar
prominences has benefited greatly in recent
years from the increased resolution of instrumentation. High
resolution observations of both filaments (on-disk)
and prominences (off-limb) give a crucial insight into the mass
maintenance and magnetic field structure that is 
needed to fully comprehend how these cool, dense structures form and
survive within the corona. 

Prominences are formed from individual
threads and knots of mass \citep{engvold76} that are suspended within
the magnetic field of the corona.  They are typically one hundred times
cooler and denser than the coronal average
\citep{labrosse10_rev_i} and are formed over polarity
inversion lines (PIL) \citep{babcock55}. The role of the magnetic
field in structuring and maintaining prominence material can be examined by
mapping the flow of plasma along field-aligned filament threads. These
flows can highlight the anatomy of the magnetic field and could be
used to test the veracity of many of the current models \citep[see
  e.g.,][for a comprehensive review]{mackay10_rev_ii}. 

The majority of observations that examine prominence threads are
taken in H$\alpha$ which typically yield spatial resolutions higher
than have been achievable in the extreme ultra-violet (EUV) and
X-ray. \cite{lin05} used
data from the Swedish Solar Telescope and found that filament threads
had an average width of 0.3'' but postulated that since this was near
the instrument's resolution, that finer structures were likely to
exist. EUV observations have shown that the higher temperature
components of filaments are wider than their H$\alpha$
counterparts due to absorption by the hydrogen Lyman
continuum \citep{heinzel01, schmieder04b}. This may have an impact on
the fundamental scale that these threads can be measured to at EUV
wavelengths. 

\cite{zirker98} first coined the term `counter-streaming' to describe
the bi-directional pattern of flows observed within filaments. These
motions \citep[see also e.g.,][]{engvold76, lin03, lin05, chae07,
  schmieder08, panasenco08} highlight a more complex picture of prominence dynamics
and can give clues to the physical scale of the fundamental prominence
threads as well as the source of the mass flows.

The counter-streaming flows examined by \cite{zirker98} were
observed in the wings of the H$\alpha$ line as both red and blue-shifts
indicating that plasma was flowing both towards and away from the
observer. A similar motion was observed perpendicular to the
line-of-sight along the threads of the prominence spine.

The work presented here details ultra-high resolution EUV observations
of counter-streaming flows along filament threads observed with NASA's
High-resolution Coronal Imager (Hi-C). It is the first observation of anti-parallel flows along
adjacent filament threads in the EUV. We make the distinction here
between the general term `counter-streaming' (which can refer to
bulk prominence mass flows as well as finer-scale motion), and the more
specific `anti-parallel' flows which we define as mass flows in
opposite directions observed along adjacent filament
threads. 

Section \ref{sec:obsdata} sets out a brief outline of the
Hi-C instrument and the data used in this study. Section
\ref{sec:results} sets out our results concerning
(i) the photospheric magnetic environment of the filament, (ii) the width of the 
prominence threads, (iii) the velocity of the plasma flows along the
threads, and  (iv) the temperature profile of these flows. Section \ref{sec:sumup}
presents the conclusions gathered from our analysis of this data-set.

\begin{figure}[ht!]
\centering
 \includegraphics[width=8cm,angle=0]{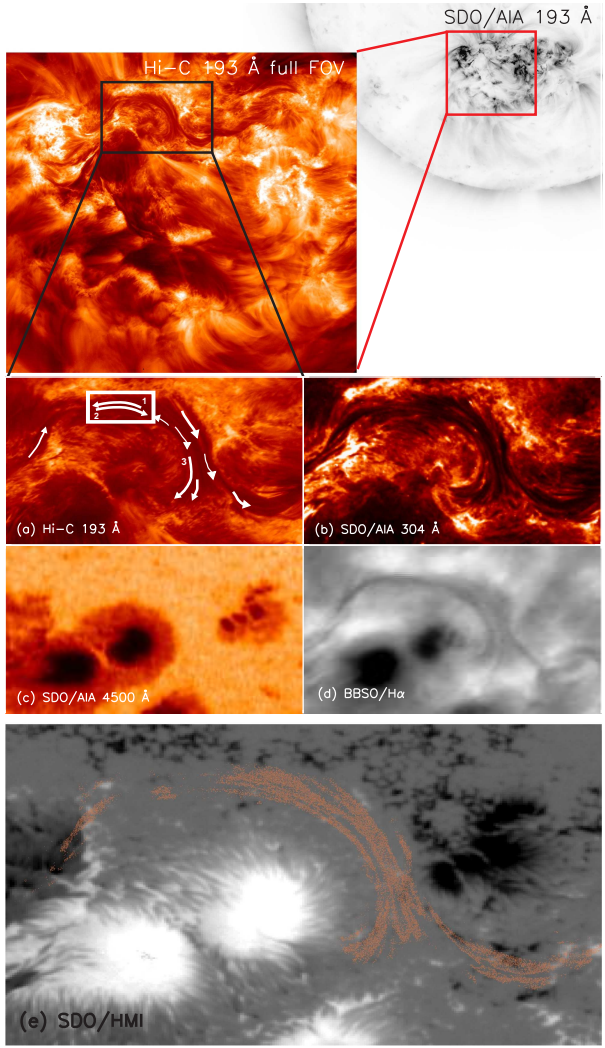}
\caption{{\it{Top right:}} partial full-disk image of the Sun taken with SDO/AIA
  193\,\AA~to show AR 11520 in context at 18:56 UT on the 11 July
  2012. The box highlights the FOV of Hi-C and the {\it{top left}} image shows an example Hi-C frame
  (at 18:55:16 UT) where the black box shows the location of the small
  filament. The four central images as well as the bottom image show
  this FOV in various co-aligned observations: {\it{(a)}} Hi-C 193\,\AA~ with the pattern of the flows
indicated by the arrows. The numbers highlight flows that were examined and the
white box shows the location of the anti-parallel flows (see online
movie XXX). {\it{(b)}}
SDO/AIA 304\,\AA\ image which shows the filament at a lower temperature
($\sim$\,50,000\,K). {\it{(c)}} SDO/AIA 4500\,\AA\ image (taken at 17:00 UT)
shows the location of the nearby sunspots in relation to the
filament. {\it{(d)}} BBSO image of the filament in H$\alpha$
(10-Jul-2012 23:30:08 UT). {\it{(e)}} SDO/HMI shows the magnetic environment of the
filament with Hi-C contours overlaid to show the position of the filament.}
 \label{fig:context}
\end{figure}

%________________________________________________________________
\section{Observations and data analysis}
\label{sec:obsdata}

The High-resolution Coronal Imager \citep[Hi-C;][]{cirtain13,kobayashi13} is a new
instrument developed by NASA's Marshall Space Flight Center and
partners\footnote{see acknowledgements} that was
launched on a sounding rocket at approximately 18:50 UT on 11 July
2012 from White Sands Missile Range. The instrument imaged a large,
magnetically-complicated active region 
(AR 11520) with the highest degree of spatial resolution ever
achieved in the EUV wavelength regime. 

The pixel
size of Hi-C is 0.1'' giving it a spatial resolution of
$\sim$\,0.3''. Comparing this with the 0.6'' pixel size and 1.2''
resolution of SDO/AIA \citep{lemen11}, Hi-C offers about four times
higher resolution. The cadence of Hi-C was $\sim$\,5.5 seconds compared to
SDO/AIA's 12 seconds. AR 11520 was imaged by Hi-C for
approximately five minutes and resulted in 200 seconds of high quality
data for the field-of-view (FOV) (7.1'\,x\,7.0') shown in Figure \ref{fig:context}
(top-left panel). 

\begin{figure}[hpb]
\centering
 \includegraphics[width=8cm,angle=0, trim= 0cm 0.5cm 0cm 0.5cm]{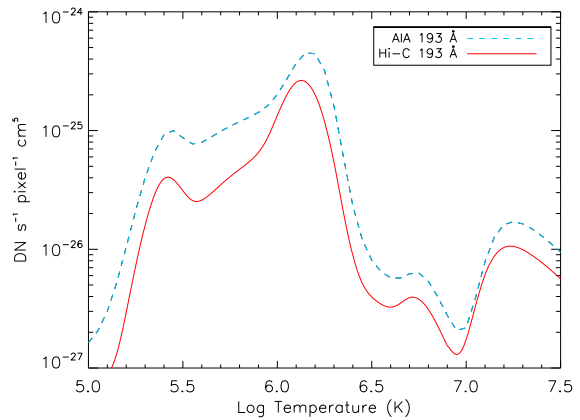}
\caption{Comparison between the temperature response functions of SDO/AIA
  193\,\AA\ and Hi-C. Note that the pixel$^{-1}$ aspect of the
  response units refers to different sizes for the two
  instruments.}
 \label{fig:tempresp}
\end{figure}

The Hi-C optics were designed to image a narrow wavelength region around
193\,\AA\ which is dominated by the Fe XII 195.119\,\AA\ line,
formed at 1.5 MK. The temperature response of Hi-C (from SSW routine HIC\_GET\_RESPONSE.PRO) is very
similar to the SDO/AIA  193\,\AA~ channel (see Figure
\ref{fig:tempresp}). The SDO/AIA temperature response function was calculated using
  the CHIANTI \citep[v7.1; ][]{landi13} `Isothermal' routine with ``coronal'' abundances, standard
  ionization equilibrium, and a density of 10$^{10}$\,cm$^{-3}$. It can
  be seen that both functions cover the same temperature regions with
  the same peaks and troughs in sensitivity.

Figure \ref{fig:context} shows the context and location of the 
filament under investigation in Hi-C as well as co-aligned data from SDO/AIA (193\,\AA,
304\,\AA, 4500\,\AA), BBSO/H$\alpha$ and SDO/HMI. Information from
additional SDO/AIA filters (94\,\AA, 131\,\AA, 171\,\AA, 193\,\AA,
211\,\AA, 335\,\AA) was also utilized in the analysis of these flows.

%________________________________________________________________

\section{Results}
\label{sec:results}

The small active region filament imaged by Hi-C exhibited clear anti-parallel flows
along adjacent threads that run perpendicular to the line-of-sight,
approximately along the spine of the filament. It is clear that these flows
are part of the filament structure as they follow the topology of the
filament channel. This is confirmed in Figure \ref{fig:context}d where the filament is
seen in H$\alpha$.  

The nature of the anti-parallel flows has been explored by
investigating the magnetic environment of the filament, the
physical width of the filament threads, the velocity of the mass flows
along these threads, and the temperature profile of the plasma that
makes up these flows.

\subsection{Magnetic structure}

AR 11520 is a magnetically complex, decaying active region. The
filament under observation is located in the middle of this region,
wrapped around a sunspot (Figure \ref{fig:context}c) and 
over a PIL (Figure \ref{fig:context}e). The filament material is
observed clearly in H$\alpha$ (Figure \ref{fig:context}d) which shows
that the filament has a two-part structure (i.e. the
main body curving around the sunspot and an additional component seen
to branch off towards the West).

The plasma flows observed with Hi-C travel along the spine of the
filament with the majority of the flows travelling Westward (shown by
the arrows in Figure \ref{fig:context}a). This suggests that the
Eastward flows may be contained within a small subset of strands along
the main filament body.   

Flows of material within filaments can give clues about the magnetic
field structure in and around these features \citep{mackay10_rev_ii}. However, caution must be
exercised as the higher density of these structures means that these
flows are not necessarily aligned with the magnetic field i.e.,
plasma is not `frozen into' the magnetic field any longer. With this
in mind, we suggest that these flows 
infer the magnetic field is horizontal along the spine of the
filament but note that this does not shed any light on the structure
of the surrounding magnetic field supporting the filament. 

The HMI photospheric
magnetogram in Figure \ref{fig:context}e shows the position of the
filament, overlaid in Hi-C contours, with respect to the PIL and
surrounding magnetic field. Along the edge of the PIL multiple
instances of magnetic flux convergence and subsequent cancellation
were observed and attributed to the interaction between the dominant polarities on
either side of the PIL, and areas of parasitic flux (i.e., smaller
portions of weaker flux with a polarity opposite to the dominant polarity
in a particular location). Various authors \citep[see e.g.,][]{martin94,
  wang07} suggest that these sites are associated with the
`footpoints' of filaments i.e., areas where the structure is rooted to
the photosphere.

Multiple footpoints could be responsible for the difference in flow
direction observed within the central part of the filament (white box
Figure \ref{fig:context}a). If the anti-parallel flows are in fact rooted in different polarity pairs,
there could be a marked difference in say, any rate of magnetic reconnection
and chromospheric evaporation taking place at each site. This in turn
could influence the direction and nature of the plasma flows. 

We interpret these flows as evidence of cool mass injection from
the chromosphere moving horizontally along the filament spine and suggest that these
anti-parallel flows are not a special-case scenario. Their particular
orientation in this active region coupled with the high spatial
resolution of Hi-C is the reason they are observed here for the first
time.

\subsection{Thread Width}

The fine-scale structure of the filament threads is an important
factor to determine. The anti-parallel flows along the filament are
 observed with Hi-C but are not as clear in
SDO/AIA. Some anti-parallel motion may be seen in SDO/AIA 193\,\AA\ and
304\,\AA\ images but it is not clearly discernible (even with
image enhancement) and we note that it would not be identifiable
without Hi-C as a guide. 

\begin{figure}[htbp]
\centering
 \includegraphics[width=8.5cm,trim=0cm 1cm 0cm 1cm]{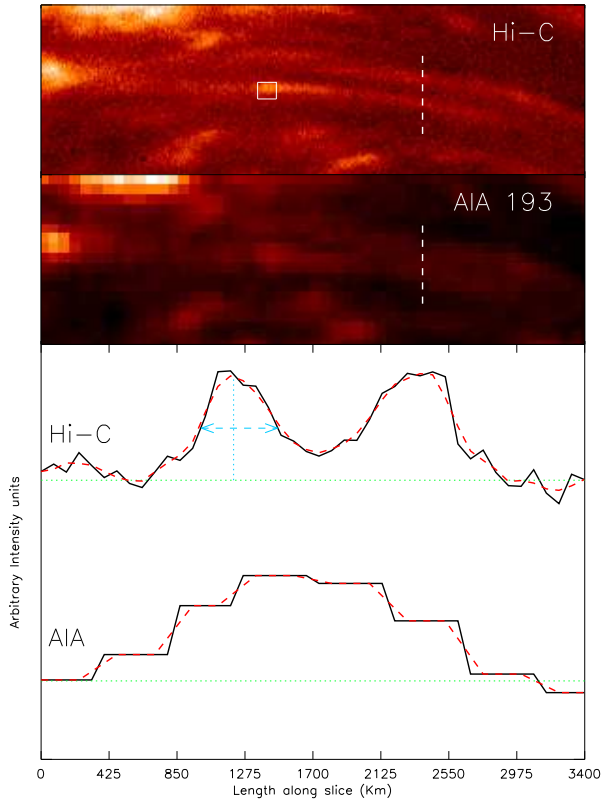}
\caption{Image showing analysis of the filament thread width. The two
  images at the top show the FOV where the anti-parallel flows are
  observed (white box Figure \ref{fig:context}a) imaged by Hi-C and
  SDO/AIA 193\,\AA. The intensity in these images was measured along the dashed white lines
  and plotted in the bottom panel. The black line in each case
  represents the intensity along this cut (averaged over 10 Hi-C
  pixels and 2 SDO/AIA pixels respectively), the dashed line (red) is the
  smoothed intensity, the dotted line (green) is the background intensity
  and the blue dashed line represents the FWHM of one of the Hi-C
  peaks which we define as the width of the thread. The white box in
  the upper image is the region used for temperature analysis in the
  next section.}
 \label{fig:structure}
\end{figure}

The width of the two adjacent threads that exhibit anti-parallel flows
is examined in Figure \ref{fig:structure}. The top portion of
this figure shows the anti-parallel flows seen by Hi-C and SDO/AIA
193\AA\ where the dashed vertical lines indicate cuts that were taken to examine the
width of the threads. It can be seen in these images that two distinct threads are
clearly distinguished by Hi-C but not by SDO/AIA. The larger pixel size of
SDO/AIA is also very apparent. The lower panel of this figure shows
the intensity profile along the cuts in each case. These profiles show
that the two threads are clearly identified by Hi-C but are merged
into one structure by the larger pixel size of SDO/AIA. 

We define the  physical width of the threads as the full width at half
maximum (FWHM) of this intensity profile (blue dashed line) and this is found to
be 0.8\,$\pm$\,0.1''. The gap between the threads was of a 
similar width (0.9''). This provides a strong indication that Hi-C has resolved
fundamentally coherent structures at this wavelength. This is further validated by the
number of Hi-C pixels contained within each thread's intensity peak (7-9) which shows
this is a clear structure well within the resolution of the
instrument. This coherence will be investigated further by examining
the speed of the flows.

\subsection{Velocity}
 
\begin{figure}[hbp]
\centering
 \includegraphics[width=8.5cm,angle=0]{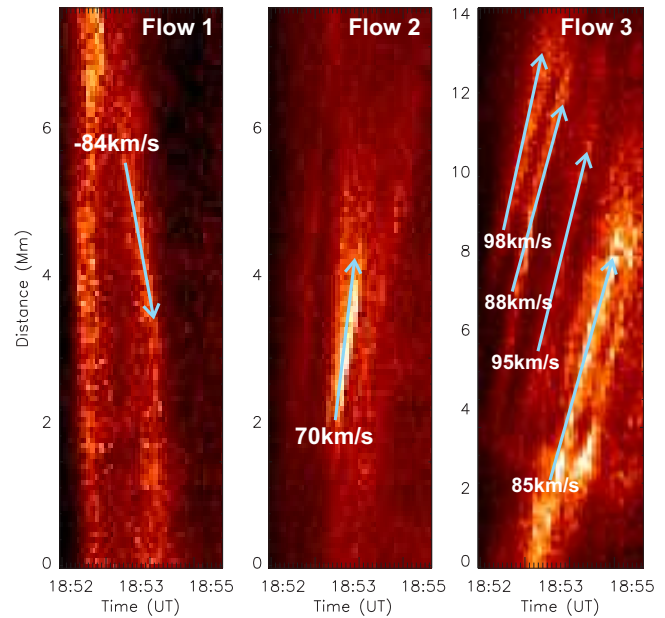}
\caption{Time-distance plots of flows seen along arrows 1, 2 and 3 in
  Figure \ref{fig:context}a. The anti-parallel flows are seen in the
  left and middle panels while multiple mass flows are seen in the right
  panel representing the flow observed to move around the edge of the sunspot.}
 \label{fig:velocity}
\end{figure}

The flows labelled 1--3 in Figure \ref{fig:context}a were
investigated further by constructing time-distance plots of intensity
 (in pixels indicated by the arrow positions) over the 200 seconds of Hi-C
data available. Additional flows were observed along the filament but were not
measurable as time-distance plots due to background noise. Figure
\ref{fig:velocity} shows these time-distance plots with the observed
flows and velocities highlighted.\\

It can be seen in Figure \ref{fig:velocity} that flows 1 and 2 are
anti-parallel and of the same order (-84.2\,$\pm$\,15.3\,km\,s$^{-1}$ and
70.2\,$\pm$\,15.6\,km\,s$^{-1}$ respectively). Flow `3' (as labelled in Figure
\ref{fig:context}a) was also examined and was seen to exhibit
multiple mass flows over the time series indicated by numerous diagonal
signatures on the time-distance plot. 

The magnitude of these
uni-directional velocity flows are of the same order as flows 1 and 2
suggesting that the anti-parallel flows observed have a typical flow
speed for this filament. Flow speeds measured by other authors in
H$\alpha$ are corresponding slower e.g., 5\,-\,20\,km\,s$^{-1}$
citep{zirker94} due to the lower temperature/higher density of plasma
imaged in this wavelength.

The larger pixel size of SDO/AIA meant that any signatures were lost
in the background intensity and thus no measurement of velocity was
possible without Hi-C.

\subsection{Temperature}

\begin{figure*}[htp]
\centering
\includegraphics[width=17cm,angle=0]{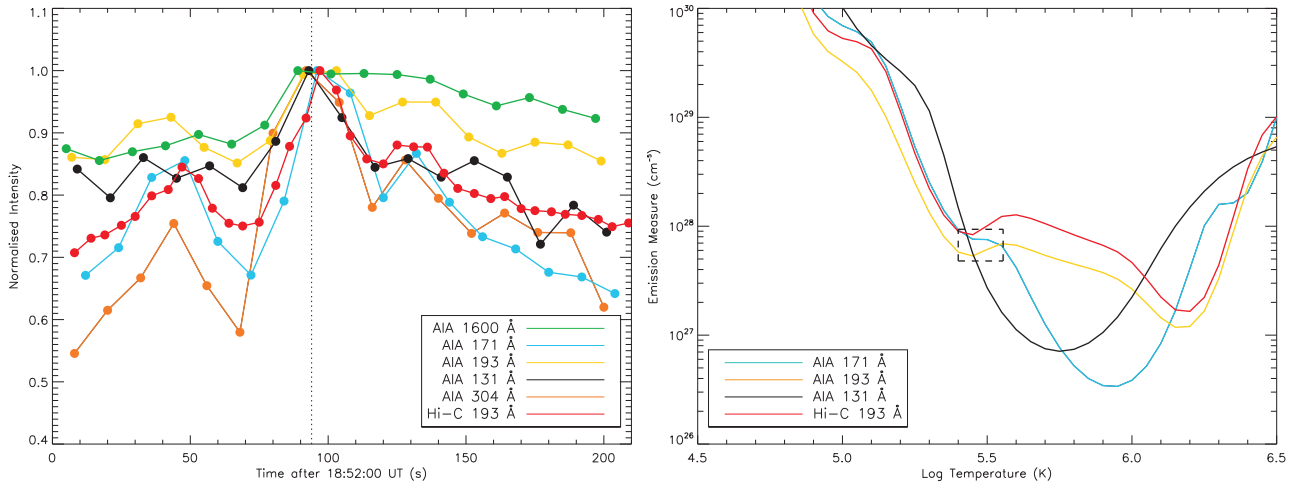}
\caption{{\it{Left:}} Normalised light-curves of the intensity change
  over time based in the area in the white box in the top panel of
  Figure \ref{fig:structure}. It can be seen that all six filters peak in the same 5\,-\,10
  second period indicated by the vertical dotted line. {\it{Right:}} EM Loci curves made using Hi-C and
  various SDO/AIA EUV passbands. The black dashed-line box indicates the area where
  the most crossings occur and indicates a plasma temperature of
  log\,T\,=\,5.45\,$\pm$\,0.1.} 
 \label{fig:temperature}
\end{figure*}

SDO/AIA provides valuable context for analysis of the Hi-C observations and allows
the active region filament to be viewed over numerous wavelengths thereby
allowing information regarding the area's temperature distribution to
be examined. Data from eight SDO/AIA channels (94\,\AA, 131\,\AA, 171\,\AA, 193\,\AA,
211\,\AA, 304\,\AA, 335\,\AA, 1600\,\AA) in the region of the anti-parallel flows was employed
to determine how the intensity of the flows changed when viewed in
different filters. A small region on one of the strands exhibiting
anti-parallel flows was chosen for
investigation (Figure \ref{fig:structure} white box). The left panel
of Figure \ref{fig:temperature} shows normalised light-curves in
five of these SDO/AIA channels (plus Hi-C) which exhibited a peak in
intensity (channels 94\,\AA\, 211\,\AA\, and 335\,\AA\ exhibited a flat intensity
profile over the time series). 

Six filters exhibited a peak in intensity at around 18:53:35 UT (dotted
line in left panel of Figure \ref{fig:temperature}) with 
the spread of the peaks being $\pm$\,10\,s around this time. Since the temporal cadence
of SDO/AIA is 12\,s, it is not possible to quantify each filter's peak
emission time with more accuracy so we shall consider these peaks
co-temporal within the instrumental constraints. 

This co-temporal intensity peak suggests that either the plasma is
very multithermal or is isothermal at a temperature that all
these filters are sensitive to. It is very likely that this is
 a mass flow as opposed to a heating event creating a conduction
front as the peaks would occur in a clear order with a time-lag in
between \citep[see e.g.,][]{viall11} if this was the case.

The temperature distribution of these flows was further investigated
using the Emission Measure Loci method \citep[see
  e.g.,][]{delzanna02}. The constructed plot 
is shown in the right panel of Figure \ref{fig:temperature} and was made using the same
intensity values gathered from the region highlighted by the small
white box in Figure \ref{fig:structure}. 

The SDO/AIA temperature response
functions were calculated using the method described by
\cite{delzanna11} (see Section \ref{sec:obsdata}). Caution must be
exercised when using imager data for this type of analysis as there
are a number of well-known problems with some of the SDO/AIA filters
e.g. cross-talk, unidentified lines, second order contributions
\citep[see e.g.,][]{dwyer10, delzanna11, boerner12}. The four filters 
used for EM loci analysis were selected as they showed a peak in intensity at this
time and are not representative of transition region plasma (as the
SDO/AIA 304\,\AA and 1600\,\AA are).

It can be seen that the four curves cross within the area
highlighted by the dashed black box. The middle of this box is found to be at
log T\,(K) = 5.45\,$\pm$\,0.1. This temperature is higher than the average temperature
of filament material which is of the order of 10$^4$\,K
\citep{labrosse10_rev_i}.

%________________________________________________________________

\section{Discussion and conclusions}
\label{sec:sumup}

This work has highlighted anti-parallel flows observed along adjacent
filament threads as seen by Hi-C. It has been shown that although SDO/AIA gives valuable
context and multi-wavelength observations, the instrument cannot
identify these small-scale flows or measure their width or
velocity. This work, as well as additional work using Hi-C \citep[see
][]{cirtain13, testa13, winebarger13, regnier13}, has
explicitly shown that Hi-C captured phenomena never seen
before and which cannot be properly examined using current instrumentation.

The pattern of flows seen along the filament suggests that the flows
are closely related to the magnetic footpoints of 
the structure. It is likely that the two threads exhibiting
anti-parallel flows are 
separate sub-structures within the filament as the coherence
of the velocity flows suggests the plasma is flowing along different
flux tubes. The flows travel horizontally along the spine of the
filament but we have no information about the inclination of the
individual threads so cannot comment on the overall structure of the
magnetic field.

Measuring the width of the filament threads established that the
structures are spatially coherent. However it is the velocity coherence
that demonstrates these anti-parallel flows are contained within
separate sub-structures. Furthermore, it is possible these
separate threads are `bundles' of smaller-scale thread-like structures
exhibiting the same pattern of mass flows. 

Based on the calculated
velocities, these anti-parallel flows are similar to other flows
observed within the filament suggesting that they are not a special
case and therefore may be ubiquitous within all filaments/prominences.   

The temperature of one of the flows is indicated to be log T\,(K) =
5.34\,$\pm$\,0.2 ($\sim$\,200,000\,K) which is supported by the
corresponding intensity enhancements seen in the cooler SDO/AIA
304~\AA\ and 1600~\AA\ filters. This temperature is around ten times
higher than the norm for filaments and could be due to the proximity
of the intense magnetic field around the sunspot, or because the plasma 
flows are the result of chromospheric evaporation. Either way, this temperature agrees with the
observations presented. 

Studying small-scale filament flows such as those observed by Hi-C
can give valuable information concerning fundamental questions such as
the source and sinks of the mass flows, as well as the orientation of
the magnetic field within these structures. From these observations it
seems the magnetic field within this filament is parallel to
the solar disk rather than helical in nature - something which can
impact a lot of the current models on prominence structure.

This is the first direct, sub-arcsecond observation of
anti-parallel/counter-streaming flows along filament threads at EUV
wavelengths and showcases the power of Hi-C's higher 
spatial and temporal resolution. Future work will expand upon this analysis by
incorporating modelling of the filament magnetic field using SDO/HMI
data. This will allow us to comment further on the origin of the
anti-parallel flows and also to extend this work to other examples of
small-scale filament flows.

%________________________________________________________________

\acknowledgements
CEA would like to thank Olga Panasenco and Giulio Del Zanna for very
useful discussions about this work. MSFC/NASA led the mission
and partners include the Smithsonian Astrophysical Observatory in
Cambridge, Mass.; Lockheed Martin's Solar Astrophysical Laboratory in
Palo Alto, Calif.; the University of Central Lancashire in Lancashire,
England; and the Lebedev Physical Institute of the Russian Academy of
Sciences in Moscow. CHIANTI is a collaborative project involving George Mason University,
the University of Michigan (USA) and the University of Cambridge
(UK).


\begin{thebibliography}{}

\bibitem[{{Babcock} \& {Babcock}(1955)}]{babcock55}
{Babcock}, H.~W. \& {Babcock}, H.~D. 1955, \apj, 121, 349

\bibitem[{{Boerner} {et~al.}(2012){Boerner}, {Edwards}, {Lemen}, {Rausch},
  {Schrijver}, {Shine}, {Shing}, {Stern}, {Tarbell}, {Title}, {Wolfson},
  {Soufli}, {Spiller}, {Gullikson}, {McKenzie}, {Windt}, {Golub}, {Podgorski},
  {Testa}, \& {Weber}}]{boerner12}
{Boerner}, P., {Edwards}, C., {Lemen}, J., {et~al.} 2012, \solphys, 275, 41

\bibitem[{{Chae} {et~al.}(2000){Chae}, {Denker}, {Spirock}, {Wang}, \&
  {Goode}}]{chae00}
{Chae}, J., {Denker}, C., {Spirock}, T.~J., {Wang}, H., \& {Goode}, P.~R. 2000,
  \solphys, 195, 333

\bibitem[{{Chae} {et~al.}(2007){Chae}, {Park}, \& {Park}}]{chae07}
{Chae}, J., {Park}, H.-M., \& {Park}, Y.-D. 2007, Journal of Korean
  Astronomical Society, 40, 67

\bibitem[{{Cirtain} {et~al.}(2013){Cirtain}, {Golub}, {Winebarger}, {De
  Pontieu}, {Kobayashi}, {Moore}, {Walsh}, {Korreck}, {Weber}, {McCauley},
  {Title}, {Kuzin}, \& {DeForest}}]{cirtain13}
{Cirtain}, J.~W., {Golub}, L., {Winebarger}, A.~W., {et~al.} 2013, \nat, 493,
  501

\bibitem[{{Del Zanna} {et~al.}(2002){Del Zanna}, {Landini}, \&
  {Mason}}]{delzanna02}
{Del Zanna}, G., {Landini}, M., \& {Mason}, H.~E. 2002, \aap, 385, 968

\bibitem[{{Del Zanna} {et~al.}(2011){Del Zanna}, {O'Dwyer}, \&
  {Mason}}]{delzanna11}
{Del Zanna}, G., {O'Dwyer}, B., \& {Mason}, H.~E. 2011, \aap, 535, A46

\bibitem[{{Deng} {et~al.}(2002){Deng}, {Lin}, {Schmieder}, \&
  {Engvold}}]{deng02}
{Deng}, Y., {Lin}, Y., {Schmieder}, B., \& {Engvold}, O. 2002, \solphys, 209,
  153

\bibitem[{{Engvold}(1976)}]{engvold76}
{Engvold}, O. 1976, \solphys, 49, 283

\bibitem[{{Heinzel} {et~al.}(2001){Heinzel}, {Schmieder}, \&
  {Tziotziou}}]{heinzel01}
{Heinzel}, P., {Schmieder}, B., \& {Tziotziou}, K. 2001, \apjl, 561, L223

\bibitem[{{Kobayashi} {et~al.}(2013){Kobayashi}, {Cirtain}, {Golub},
  {Winebarger}, {De Pontieu}, {Moore}, {Walsh}, {Korreck}, {Weber}, {McCauley},
  {Title}, {Kuzin}, \& {DeForest}}]{kobayashi13}
{Kobayashi}, K., {Cirtain}, J.~W., {Golub}, L., {et~al.} 2013, xx

\bibitem[{{Labrosse} {et~al.}(2010){Labrosse}, {Heinzel}, {Vial}, {Kucera},
  {Parenti}, {Gun{\'a}r}, {Schmieder}, \& {Kilper}}]{labrosse10_rev_i}
{Labrosse}, N., {Heinzel}, P., {Vial}, J.-C., {et~al.} 2010, \ssr, 151, 243

\bibitem[{{Landi} {et~al.}(2013){Landi}, {Young}, {Dere}, {Del Zanna}, \&
  {Mason}}]{landi13}
{Landi}, E., {Young}, P.~R., {Dere}, K.~P., {Del Zanna}, G., \& {Mason}, H.~E.
  2013, \apj, 763, 86

\bibitem[{{Lemen} {et~al.}(2012){Lemen}, {Title}, {Akin}, {Boerner}, {Chou},
  {Drake}, {Duncan}, {Edwards}, {Friedlaender}, {Heyman}, {Hurlburt}, {Katz},
  {Kushner}, {Levay}, {Lindgren}, {Mathur}, {McFeaters}, {Mitchell}, {Rehse},
  {Schrijver}, {Springer}, {Stern}, {Tarbell}, {Wuelser}, {Wolfson}, {Yanari},
  {Bookbinder}, {Cheimets}, {Caldwell}, {Deluca}, {Gates}, {Golub}, {Park},
  {Podgorski}, {Bush}, {Scherrer}, {Gummin}, {Smith}, {Auker}, {Jerram},
  {Pool}, {Soufli}, {Windt}, {Beardsley}, {Clapp}, {Lang}, \&
  {Waltham}}]{lemen11}
{Lemen}, J.~R., {Title}, A.~M., {Akin}, D.~J., {et~al.} 2012, \solphys, 275, 17

\bibitem[{{Lin} {et~al.}(2005){Lin}, {Engvold}, {Rouppe van der Voort}, {Wiik},
  \& {Berger}}]{lin05}
{Lin}, Y., {Engvold}, O., {Rouppe van der Voort}, L., {Wiik}, J.~E., \&
  {Berger}, T.~E. 2005, \solphys, 226, 239

\bibitem[{{Lin} {et~al.}(2003){Lin}, {Engvold}, \& {Wiik}}]{lin03}
{Lin}, Y., {Engvold}, O.~R., \& {Wiik}, J.~E. 2003, \solphys, 216, 109

\bibitem[{{Mackay} {et~al.}(2010){Mackay}, {Karpen}, {Ballester}, {Schmieder},
  \& {Aulanier}}]{mackay10_rev_ii}
{Mackay}, D.~H., {Karpen}, J.~T., {Ballester}, J.~L., {Schmieder}, B., \&
  {Aulanier}, G. 2010, \ssr, 151, 333

\bibitem[{{Martin} \& {Echols}(1994)}]{martin94}
{Martin}, S.~F. \& {Echols}, C.~R. 1994, in Solar Surface Magnetism, ed. R.~J.
  {Rutten} \& C.~J. {Schrijver}, 339

\bibitem[{{O'Dwyer} {et~al.}(2010){O'Dwyer}, {Del Zanna}, {Mason}, {Weber}, \&
  {Tripathi}}]{dwyer10}
{O'Dwyer}, B., {Del Zanna}, G., {Mason}, H.~E., {Weber}, M.~A., \& {Tripathi},
  D. 2010, \aap, 521, A21

\bibitem[{{Panasenco} \& {Martin}(2008)}]{panasenco08}
{Panasenco}, O. \& {Martin}, S.~F. 2008, ASP Conf. Ser., 383, 243

\bibitem[R{\'e}gnier et al.(2013)]{regnier13}
        {R{\'e}gnier}, S., {Alexander}, C.~E., {Walsh}, R.~W., {Cirtain},
J.~W., {Winebarger},  A.~R., {Golub}, L., {Korreck}, K.~E.,
{DePontieu}, B., {Kobayashi}, K.,  {Korreck}, K.~E., {Mitchell}, N.,
{Platt}, S., {Weber}, M., {De Pontieu}, B., {Title}, A., {Kobayashi},
K., {Kuzin}, S., {DeForest}, C.~E. 2013, in prep.



\bibitem[{{Schmieder} {et~al.}(2008){Schmieder}, {Bommier}, {Kitai},
  {Matsumoto}, {Ishii}, {Hagino}, {Li}, \& {Golub}}]{schmieder08}
{Schmieder}, B., {Bommier}, V., {Kitai}, R., {et~al.} 2008, \solphys, 247, 321

\bibitem[{{Schmieder} {et~al.}(2004){Schmieder}, {Lin}, {Heinzel}, \&
  {Schwartz}}]{schmieder04b}
{Schmieder}, B., {Lin}, Y., {Heinzel}, P., \& {Schwartz}, P. 2004, \solphys,
  221, 297

\bibitem[{{Testa} {et~al.}(2013){Testa}, {De Pontieu},
  {Mart{\'{\i}}nez-Sykora}, {DeLuca}, {Hansteen}, {Cirtain}, {Winebarger},
  {Golub}, {Kobayashi}, {Korreck}, {Kuzin}, {Walsh}, {DeForest}, {Title}, \&
  {Weber}}]{testa13}
{Testa}, P., {De Pontieu}, B., {Mart{\'{\i}}nez-Sykora}, J., {et~al.} 2013,
  \apjl, 770, L1

\bibitem[{{Viall} \& {Klimchuk}(2011)}]{viall11}
{Viall}, N.~M. \& {Klimchuk}, J.~A. 2011, \apj, 738, 24

\bibitem[{{Wang} \& {Muglach}(2007)}]{wang07}
{Wang}, Y.-M. \& {Muglach}, K. 2007, \apj, 666, 1284

\bibitem[Winebarger et al.(2013)]{winebarger13}{Winebarger}, A.~R., {Cirtain}, J.~W., {Golub}, L., {DePontieu}, B.,
	{Kobayashi}, K., {Moore}, R.~L., {Walsh}, R.~W., {Korreck}, K.~E.,
	{McCauley}, P., {Title}, A., {Kuzin}, S., \& {DeForest},
        C.~E. 2013, submitted
	


\bibitem[{{Zirker} {et~al.}(1998){Zirker}, {Engvold}, \& {Martin}}]{zirker98}
{Zirker}, J.~B., {Engvold}, O., \& {Martin}, S.~F. 1998, \nat, 396, 440

\end{thebibliography}
\end{document}